\documentclass[12pt, onecolumn]{IEEEtran}
\usepackage{graphicx} 
\usepackage{xcolor}
\usepackage{float}
\usepackage{newtxtext}
\usepackage{mathtools, bm, bbm, mathrsfs, amssymb}
\usepackage{url, hyperref, cite}
\usepackage{setspace}
\def\BibTeX{{\rm B\kern-.05em{\sc i\kern-.025em b}\kern-.08em
\DeclareMathOperator*{\argmax}{arg\,max}
    T\kern-.1667em\lower.7ex\hbox{E}\kern-.125emX}}
 \AtBeginDocument{\definecolor{ojcolor}{cmyk}{0.93,0.59,0.15,0.02}}
 
\DeclareMathOperator*{\argmax}{\textsf{arg max}}
\newcommand{\E}{\mathbbm{E}}
\newcommand{\SNR}{\textsf{SNR}}
\newcommand{\frc}{f_{\textsf{RC}}}
\newcommand{\mC}{\mathcal{C}}
\newcommand{\mA}{\mathcal{A}}
\newcommand{\mW}{\mathcal{W}}
\newcommand{\sigmath}{\sigma_{\textsf{th}}}
\newcommand{\sigmash}{\sigma_{\textsf{sh}}}
\newcommand{\mR}{\mathcal{R}}
\newcommand{\lambdasm}{\Lambda_s^{(m)}}
\newcommand{\fh}{f_{\mathbbm{h}}(h)}

\begin{document}
\title{Optimal Photodetector Size for High-Speed Free-Space Optics Receivers}
\author{Muhammad~Salman~Bashir,~\emph{Senior~Member,~IEEE}, Qasim~Zeeshan~Ahmed,~\emph{Member,~IEEE}, and Mohamed-Slim~Alouini,~\emph{Fellow, IEEE}
\thanks{Muhammad Salman Bashir and Qasim Zeeshan Ahmed are with the School of Computing and Engineering, University of Huddersfield, HD1 3DH, England, United Kingdom. }
\thanks{Mohamed-Slim Alouini is with Computer, Electrical
and Mathematical Sciences and Engineering (CEMSE) Division, King Abdullah University of Science and Technology (KAUST), Thuwal, 23955-6900, Saudi Arabia.} 
\thanks{email: (m.bashir@hud.ac.uk; q.ahmed@hud.ac.uk; slim.alouini@kaust.edu.sa)}.}
\maketitle

\begin{abstract}
    The selection of an optimal photodetector area is closely linked to the attainment of higher data rates in optical wireless communication receivers. If the photodetector area is too large, the channel capacity degrades due to lower modulation bandwidth of the detector. A smaller photodetector maximizes the bandwidth, but minimizes the captured signal power and the subsequent signal-to-noise ratio. Therein lies an opportunity in this trade-off to maximize the channel rate by choosing the optimal photodetector area.   In this study, we have optimized the photodetector area in order to maximize the channel capacity of a free-space  optical link for a diverse set of communication scenarios.  We believe that the study in this paper in general---and the closed-form solutions derived in this study in particular---will be helpful to maximize achievable data rates of a wide gamut of optical wireless communication systems: from long range deep space optical links to short range indoor visible light communication systems. 
\end{abstract}


\section{Introduction}

Due to the availability of large unregulated spectrum in the optical domain of electromagnetic waves, free-space optics (FSO)---also commonly known as optical wireless or laser communications---is an important candidate for supporting high data-rates in the sixth generation (6G) and beyond wireless networks \cite{Trichili:JOSAB:20}. FSO has already been a major player in non-terrestrial networks such as high data-rate inter-satellite communications in low-Earth orbit, medium-Earth orbit, geosynchronous-equatorial orbit, and high-Earth orbit. FSO has also found extensive deployment in integrated space-to-ground and ground-to-space networks \cite{Hemmati:JPROC:11}. The National Aeronautics and Space Administration (NASA) successfully executed downlink data rates up to 622 Mbps for a Moon-to-Earth link in the Aerospace Corporation’s Optical Communication and Sensor Demonstration (OCSD) in 2017 \cite{Goorjian:SPIE:21}. Additionally, NASA's TeraByte InfraRed Delivery (TBIRD) system promises data rates of more than  200 Gbps  for a CubeSat-to-ground link using FSO communications \cite{Chang:MIT:19, Goorjian:SPIE:21}. Due to narrow beamwidths associated with optical signals, FSO is used exclusively for deep space communications for link lengths that may stretch over a few hundred thousand kilometers \cite{Hemmati:JPROC:11}. Recently, scientists at ETH Z\"{u}rich were able to achieve 1 Tbps speed with a single wavelength free-space laser link that spanned $53$ kilometers from the Jungfraujoch to Bern in Switzerland \cite{Horst:Light_Sci_App:23}.

\subsection{Trade-off Between Detector Bandwidth and SNR}
Laser photodiodes are the basic detection elements in a modern optical receiver that are used not only for detection of optical symbols but also serve as position or angle-of-arrival sensors in a quad detector array configuration \cite{Snyder1991, Bashir:TCOMM:a:21}. Photodiodes are classified as either P-N, P-I-N or avalanche photodiodes \cite{Pearsall:MH:09}. As the name implies, a P-N photodiode is made up of a P-N junction---a boundary between two types of semiconductor materials, the P-type semiconductor and N-type semiconductor. A  P-I-N diode consists of a undoped intrinsic semiconductor region between the P-type and the N-type semiconductor materials. P-I-N photodiodes have faster switching capability compared to their P-N counterparts \cite{Haberlin:Wiley:12}. Avalanche photodiodes (APD) are either based on Silicon, Germanium or InGaAs semiconducting materials. A distinguishing feature of APD's is that they are operated with a high reverse voltage that allows them to detect very low levels of signal energy. This feature endows them with high sensitivity and high signal-to-noise ratio characteristics compared to P-N and P-I-N photodiodes \cite{MCIntyre:T-ED:66, Hayat:JQE:92}.

The existence of P-N junction in a photodiode leads to  capacitive $C$, and resistive $R$ effects that can limit the switching rate---or the response speed---of a photodiode. This phenomenon is captured by $RC$ time constant of a photodiode: a larger $RC$ time constant leads to a reduced response time or modulation frequency (or cut-off frequency) of a photodiode and vice versa \cite{AMRAOUI:optlastec:22}. In order to be able to transmit high-data rates in optical wireless channels, a low $RC$ time constant or a high modulation frequency is highly desirable. Since the capacitance of a diode is directly related to its active area, photodiodes with a smaller area lead to a higher modulation frequency or bandwidth.  On the other hand, if the active area of the photodiode is too small, the signal energy captured by the photodiode diminishes. This leads to poor signal-to-noise ratio at the receiver photodiode and a poor signal detection performance. Thus, a trade-off exists between diode bandwidth and signal-to-noise ratio, and we ought to choose the optimal diode size or area in order to maximize the communication performance of the optical receiver.

\subsection{Motivation of Current Study} \label{motivate}

\begin{figure}
    \centering
    \includegraphics[scale=0.62]{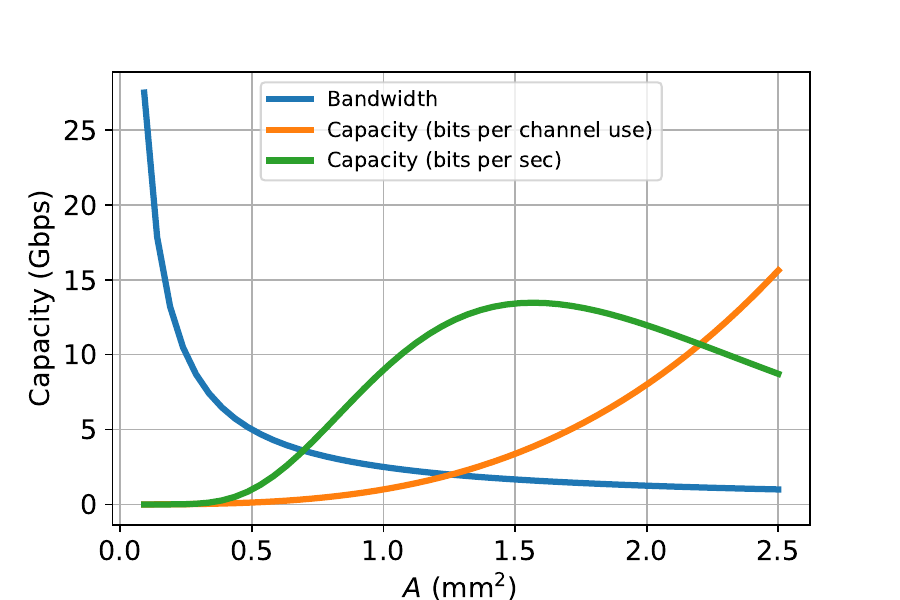}
    \caption{Channel capacity as a function of detector area.}
    \label{fig5}
\end{figure}

The channel capacity is a useful metric that captures both the bandwidth as well as the signal-to-noise ratio of a communication channel. This metric was first employed in \cite{Azarkh:SPIE:23} to illustrate that the throughput (or capacity) of an pulse amplitude modulation optical wireless channel is maximized at a particular photodetector area. In order to demonstrate this concept,  the channel capacity of a coherent Gaussian optical wireless channel---denoted by $\mC$---is furnished by the well-known \emph{Shannon-Hartley} theorem \cite{Chaaban:COMST:22}:
\begin{align}
    \mC& = W \mC_0, \nonumber \\
    & = W \log_2 \left( 1 + \frac{\bar{P} }{ N_0 W  }\right), \label{sigmath0}
\end{align}
where $W$ is the bandwidth in Hz, $\mC_0$ is channel capacity in bits per channel use, $\bar{P}$ is the average received signal power in Watts and $N_0$ is the two-sided power spectral density in Watts/Hz. Here, the quantity $\frac{\bar{P}}{N_0W}$ is the signal-to-noise ratio. In \eqref{sigmath0}, we observe that the Gaussian channel capacity (measured in bits per second)  is the product of two terms---the bandwidth $W$ and the channel capacity in terms of bits per channel use, $\mC_0$. We also observe that the signal-to-noise ratio in $\mC_0$ is itself inversely proportional to the bandwidth $W$. The bandwidth $W$ is inversely proportional to the area $A$ of the photodiode, whereas signal-to-noise ratio is directly proportional to the cube of photodiode active area: $A^3$. Thus, for a very small $A$, the available modulation bandwidth will be quite large, whereas the signal-to-noise ratio---and, therefore, the capacity $\mC_0$---will diminish. On the other hand, a large photodiode area $A$ leads to high capacity per channel use $\mC_0$ but negligible bandwidth $W$. This hints at the existence of an optimal trade-off---in terms of best or optimal area $A$---between bandwidth $W$ and  $\mC_0$ that leads to the  maximization of the product (of these two terms): the channel capacity in bits per second. This phenomenon is illustrated in Fig.~\ref{fig5} where the bandwidth $W$ is monotonically decreasing in $A$, the channel capacity $\mC_0$ (bits per channel use) is increasing monotonically with area $A$, and the product (channel capacity in bits per seconds) is maximized at a certain (optimal) area $A$.

Even though the study \cite{Azarkh:SPIE:23} hints  at the existence of an optimal photodetector area for maximization of channel capacity through a graphical illustration, they did not provide any closed-form analytical solutions for the optimal photodetector area. Moreover, the said study considers only one type of channel: the Gaussian laser channel that is not affected by signal fading.  Since the optimization of photodetector area is an important problem in our estimation, there is a strong motivation  to look for optimal (or even suboptimal) closed-form analytical solutions of the photodetector area for the maximization of Gaussian channel capacity. Additionally, signal-dependent shot noise appears in optical systems where the effect of background radiation can be quite large, and this requires optimization of photodetector area for signal dependent Gaussian noise channels. Finally, since detector arrays have been shown to provide significant performance improvement over a single detector for optical symbol detection \cite{Tsai:OJCOMS:22}, maximization of capacity of channels involving detector array receivers---as a function of individual detector size---is an important problem for future optical wireless communication systems.  All of the above mentioned considerations form the motivation of the current study that deals with the optimization of photodetector area for a diverse set of free-space optical communication scenarios.

\subsection{Background Literature Review }

In this paper, we examine the optimization of the photodetector area to maximize the channel capacity of an optical wireless channel. For this problem, we have classified the relevant literature review under four categories: i) Relationship between modulation bandwidth of a photodiode  and its area, ii) signal dependent Gaussian noise models for optical wireless channels, iii) pointing error/atmospheric turbulence-induced fading in optical wireless channels and iv) optical wireless communications with a detector array receiver.


There are a number of studies in current literature that discuss the relationship between detector area and modulation bandwidth. For instance, the paper \cite{AMRAOUI:optlastec:22} proposes a novel P-I-N photodiode based on InGaAsN lattice which is matched to GaAs. The proposed photodiode consists of double transparent layers for the depletion region and performs high speed photodetection compared to a regular P-I-N photodiodes. The authors in \cite{ELBESSEGHI:matchemphys:15} consider frequency response optimization of P-I-N photodiodes based on GaN/InGaN semiconducting materials that operate on $633$~nm wavelength. In the said study, the authors examine frequency response optimization of the diode by using mixed depletion region rather than a single absorbing depletion region. 
The study \cite{Xu:JLT:09} demonstrates design considerations and simulation aspects for multiple cladding layer GaAs-based P-I-N waveguide photodiodes. The authors in this study demonstrated that the theoretical cut-off frequency limit of $80$~GHz can be achieved by using a thin absorption layer of $0.4~\mu$m. In \cite{Alshehri:JPHOT:17}, the authors consider design and fabrication of high speed InGaN/GaN based photodiodes by reducing the capacitive effects. The authors in this study demonstrated a cut-off frequency of $300$~MHz using Noise measurement for $100 \times 100~\mu$m$^2$ photodiodes. According to the authors, this result will enable a signficantly better higher frequency response for photodiode dimensions up to $20~\mu$m.

We now consider some important  references on the evaluation of channel capacity  as well as noise models for free-space optical links. The authors in \cite{Lapidoth:TIT:09} derived several upper and lower bounds for a noncoherent intensity modulated direct detection free-space optical link for constrained average and peak power. Farid et al. \cite{Farid:JSAC:09} consider the design of nonuniform optical intensity signaling that approaches capacity for constrained average and peak signal amplitudes. In their work, they derive a simple expression for capacity-approaching distribution via source entropy maximization. The study \cite{Wang:TWC:20} derives tight capacity bounds for indoor visible light communication channels in the presence of signal dependent noise whereas the articles  \cite{Moser:TIT:12, Khan:AUSCTW:11} considers various capacity bounds for a free-space optical channel in the presence of signal dependent Gaussian noise. 

There are a number of studies on the modeling of pointing errors in free-space optical communications. The authors in \cite{Quwaiee:TWC:16, Issaid:TWC:17} consider generalized models of pointing error for capacity calculations, whereas \cite{Ansari:TWC:16, Zedini:TWC:17} consider pointing error modeling along with turbulence induced fading in FSO links. The authors in \cite{Bashir:TCOMM:22, Bashir:TWC:22} exclusively consider Rayleigh distribution-based pointing error model for dual and multihop relaying FSO channels that involve hovering unmanned aerial (UAV) based relays. The study \cite{Farid:JLT:07} considers maximization of outage capacity of an optical wireless link subject to pointing errors. Tsai et al. \cite{Tsai:OJCOMS:22} examine the effect of pointing error on the performance of a detector array receiver in a free-space optical link. 

We wrap up the literature review section with the mention of some references on detector array receivers in free-space optical communications. The authors consider detector arrays for joint beam tracking and symbol detection in \cite{Bashir:TAES:16, Bashir:TAES:17, Bashir:TAES:18, Bashir:OJCOMS:21}, whereas the studies \cite{Bashir:TWC:20, Bashir:TCOMM:21} consider FSO acquisition algorithms based on detector array receiver. The article \cite{Bashir:TAES:19} consider time synchronization of pulse position modulation signals based on an array of detectors. The study \cite{Bashir:TAES:20} examines the effect of the number of array elements on the performance of the maximal ratio combiner that is used to fuse the outputs of the array detectors. The article \cite{Bashir:OJCOMS:20} explores multiple-input-single-output schemes for a detector array receiver in free-space optical communications. 

\subsection{Contributions of Our Study}

\begin{figure}
    \centering
    \includegraphics[scale=0.6]{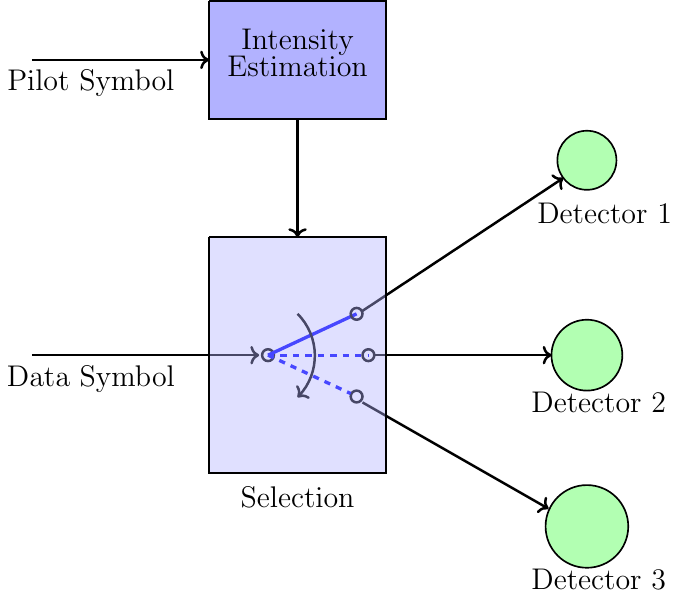}
    \caption{Block diagram of adaptive optical receiver.}
    \label{adaptive_receiver}
\end{figure}

{With the exception of \cite{Azarkh:SPIE:23} and our understanding, there is no exiting research or literature that analyzes the relationship between channel capacity and the photodetector area of an optical wireless communication channel.} Even though the said paper highlights the fact that the channel capacity peaks at a certain (optimal) photodetector area, a detailed analysis and derivations of (even approximate) closed-form solutions are not provided. Additionally, the focus of \cite{Azarkh:SPIE:23} is on just one type of channel: the signal independent thermal noise channel that is modeled by a Gaussian distribution. In this study, we carried out a comprehensive analysis on the optimization of photodetector area for channel capacity maximization. Specifically,  we i) derived closed-form solution (in terms of Lambert W function) of  optimal photodetector area for the signal independent Gaussian noise channel as furnished by \eqref{opt_A}, ii) provided analytical closed-form solutions of close-to-optimal area for intensity fading channels as given by \eqref{pointing_error_soln} and \eqref{exp_turbulence_soln},  iii) analyzed channels based on a detector array receiver and provided closed-form optimal solutions for equal gain combining and maximal ratio combining arrays as supplied by \eqref{opt_A_egc} and \eqref{opt_A_mrc}, respectively, and  iv) provided numerical solutions of optimal detector area for shot noise limited regime that is modeled by signal dependent Gaussian noise. A major conclusion of our study states that---for the signal independent Gaussian noise case---the optimal area is proportional to the factor $\left( \frac{\alpha}{\beta} \right)^{\frac{1}{m}}$ where the term  $\frac{\alpha}{\beta}$ is the inverse of the normalized signal-to-noise ratio of the optical receiver. The constant $m=3$ for channels involving either the single photodetector or the maximal ratio combining array, and $m=2$ for equal gain combining array. This implies that as the signal-to-noise ratio improves, the optimal detector area shrinks in order to exploit the extra bandwidth that comes with a smaller area. {Based on this discussion,  we propose an adaptive optical wireless communications receiver that 
 consists of a number of photodetectors with varying active areas.  {For such receivers, the ``optimal'' photodetector based on the value of estimated signal intensity is chosen at any given time in order to maximize the channel capacity.} A conceptual example of such an adaptive receiver is shown in Fig.~\ref{adaptive_receiver} which consists of a signal intensity estimation block, a selection block and three photodetectors of different areas. The intensity estimation block operates on pilot symbols to estimate signal intensity with the help of an array of detectors \cite{Bashir:OJCOMS:21}. The intensity estimate is then fed into a selection block that selects the best photodetector to maximize channel capacity for incoming data symbols.  This adaptive receiver can effectively mitigate the effect of slow intensity fading due to atmospheric turbulence or pointing error.} 

\subsection{Paper Organization}
This paper is organized as follows. Section~\ref{analysis} highlights the relationship between detector area and  bandwidth of an optical photodetector, and  Section~\ref{analysis: thermal noise} deals with the derivation and maximization channel capacity of a thermal noise limited receiver. Section~\ref{fading_section} considers the  photodetector area optimization for signal intensity fading channels. In this section,  two scenarios are considered: one  that deals with fading due to pointing error and the other that examines signal fading due to strong atmospheric turbulence that is modeled by negative exponential distribution. Section~\ref{detector array} deals with channel capacity maximization of a detector array receiver for two combining schemes: the \emph{equal gain combining} (EGC) and \emph{maximal ratio combining} (MRC). Section~\ref{analysis: general case} deals with the general case where both the signal independent thermal noise and signal dependent shot noise due to signal and background radiation are considered.    Section~\ref{experiments} is dedicated to the interpretation and commentary of experimental results, and the last section, Section~\ref{conclusion}, sums up the important results of this study.

\section{Relationship Between Detector Size and Bandwidth } \label{analysis}

\begin{figure}
    \centering
    \includegraphics[scale=0.8]{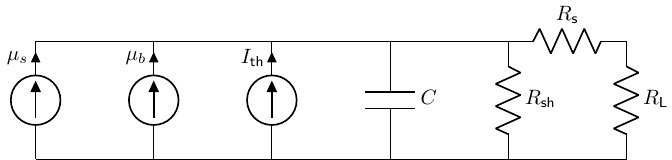}
    \caption{Equivalent circuit representation of a photodiode.}
    \label{circuit}
\end{figure}

\subsection{Detector Size and Bandwidth} \label{analysis: bandwidth}
{Fig.~\ref{circuit} provides an equivalent circuit representation of a photodiode operated in the photoconductive mode to detect light signal \cite{OSI_optoelectronics}}. In this circuit, the photodiode behaves as a photocontrolled current source in parallel with a semiconductor diode. The circuit shows three current sources: the current due to incident signal light denoted by $\mu_s$, current $\mu_b$ which is due to (unwanted) background radiation or leakage current and the current $I_{\textsf{th}}$ which is due to thermal noise in photodiode resistive contacts. The sources $\mu_b$ and $I_{\textsf{th}}$ represent noise sources in the circuit. The noise due to $\mu_b$ is the Poisson shot noise whose variance depends on the value of $\mu_b$. The noise due to $I_{\textsf{th}}$ is modeled by a zero-mean Gaussian noise process with variance $\sigma_{\textsf{th}}^2$.

The semiconductor diode in Fig.~\ref{circuit} is represented by its constituent elements: the capacitance $C$ that represents the junction capacitance of the photodiode, the photodiode shunt resistance $R_{\textsf{sh}}$ and the resistance $R_{\textsf{s}}$ generated by photodiode contacts, wire bonds and semiconductor material. The resistance $R_{\textsf{s}}$ is typically a few tens of ohms. For silicon photodiodes, the resistance $R_{\textsf{sh}}$ is usually in hundreds or thousands of megaohms, and indium gallium arsenide based photodiodes have typically even higher shunt resistance values. Therefore, in most practical scenarios, the shunt resistance $R_{\textsf{sh}}$ can be safely ignored \cite{OSI_optoelectronics}.

The $3$-dB bandwidth (or cutoff frequency) of a photodiode---denoted by $f_c$---is given by \cite{AMRAOUI:optlastec:22, ELBESSEGHI:matchemphys:15}
\begin{align}
    f_c = \frac{f_t \frc}{\sqrt{ f_t^2 + \frc^2 }},
\end{align}
where $f_t$ is the \emph{transit-time} cutoff frequency and $\frc$ is the \emph{capacitive} cut-off frequency. The transit-time limited bandwidth is determined by the electron and hole transit times through the depletion region. The transit-time limited bandwidth is expressed by the equation \cite{Xu:JLT:09}
\begin{align}
    f_t \approx \frac{0.45 \bar{\nu}}{ d},
\end{align}
where $\bar{\nu}$ is the mean carrier drift velocity and $d$ is the thickness of depletion region. Since the holes' mobility is quite limited compared to free electrons, it is typically the holes that determine the transit-time limited bandwidth \cite{Beling:JSTQE:14}. 

The capacitive cut-off frequency is given by \cite{Alshehri:JPHOT:17}
\begin{align}
    \frc = \frac{1}{2 \pi R C}, \label{frc}
\end{align}
where $R$ is the sum of photodiode series resistance and load resistance, i.e., $R = R_{\textsf{s}} + R_{\textsf{L}}$, and $C$ is the photodiode junction capacitance \cite{OSI_optoelectronics}. The junction capacitance is given by 
\begin{align}
C = \dfrac{\epsilon_0 \epsilon_r A}{d}, \label{C}
\end{align}
where $A$ is the active area of photodiode, $d$ is the depletion region thickness, $\epsilon_0$ is the permittivity of free space and $\epsilon_r$ is the relative permittivity of photodiode depletion region \cite{AMRAOUI:optlastec:22}. By combining \eqref{C} and \eqref{frc}, we have that
\begin{align}
    \frc = \frac{d}{2 \pi \epsilon_0 \epsilon_r R A}=\frac{\alpha}{A}, \label{frc1}
\end{align}
where 
\begin{align}
\alpha \coloneqq \frac{d}{2 \pi \epsilon_0 \epsilon_r R}. \label{alpha1}
\end{align}
 It has been demonstrated in \cite{AMRAOUI:optlastec:22} that for detector areas greater than $100~\mu m^2$, the capacitive effect dominates the transit-time cutoff. In this case, the overall cut-off frequency $f_c$ is majorly determined by $\frc$. Thus, when $\frc \ll f_t$ for $A > 100 \, \, \mu m^2$, we have that
\begin{align}
    f_c = \frac{f_t \frc}{\sqrt{ f_t^2 + \frc^2 }} \approx  \frac{f_t \frc}{\sqrt{ f_t^2 }} = \frc. 
\end{align}

\section{Channel Capacity Maximization of Thermal Noise Limited Receiver}\label{analysis: thermal noise}

\subsection{Signal Model}
\begin{figure}
    \centering
    \includegraphics[scale=1.2]{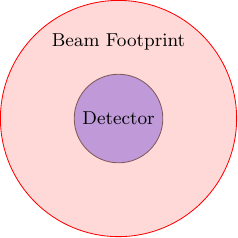}
    \caption{ {Gaussian beam footprint projected onto a circular detector in the focal plane.}}
    \label{Fig1}
\end{figure}

The signal power captured by the photodetector is given by
\begin{align}
    \mu_s& \coloneqq  \iint_{\mathcal{D}} \Lambda_s(x, y) \, dy \,  dx, \label{mu_s11}
    \end{align}
    where $\Lambda_s$ is the signal intensity measured in Watts/m$^2$ and  $\mathcal{D}$ is the region of the circular detector. The intensity in focal plane is modeled by a two-dimensional circularly symmetric Gaussian distribution \cite{Snyder1991}:
    \begin{align}
   \Lambda_s & \coloneqq  \frac{I_0}{2\pi \rho^2} \exp \left( - \frac{(x-x_0)^2 + (y-y_0)^2}{2 \rho^2} \right). \label{intensity_def}
\end{align}
 Here, $(x_0, y_0)$ is the center of the beam, $\rho$  is the beam or spot radius and $I_0$ is the total power in the beam measured in Watts. {Fig.~\ref{Fig1} shows the Gaussian beam impinging on a circular photodetector. } 
We assume that the detector is placed at the origin and that the center of the beam and the center of the detector coincide\footnote{This condition corresponds to the condition of absence of pointing error in the channel and maximum power capture by the detector.}. For $A \ll \pi \rho^2$, we have that
\begin{align}
    \mu_s \approx \mu_0 A,
\end{align}
where \begin{align}
\mu_0 \coloneqq \frac{I_0}{2\pi \rho^2} \label{peak_intensity1}
\end{align} 
is the peak intensity projected onto the focal plane.

\subsection{Channel Capacity}
If we assume that the only major source of noise in the circuit is the  \emph{thermal noise} with variance $\sigmath^2$, we have that  $\sigmath^2 = N_0 W$ where $W$ is the photodiode bandwidth  and $N_0$ is one-sided noise spectral density measured in W/Hz. Here, we assume  that the photodiode resistance is negligible compared to circuit resistance, and therefore $\sigmath^2$ is generated mainly by the circuit resistive elements. 

The Gaussian channel capacity of a coherent optical channel is reproduced below as 
\begin{align}
    \mC = \mathcal{W} \log_2 \left( 1 + \frac{\mu_s^2}{ \sigmath^2} \right), \label{capacity}
\end{align}
where the bandwidth 
\begin{align}
\mathcal{W} \coloneqq  \frc = \frac{\alpha}{A}.
\end{align}
 By substituting the expression for $\mu_s$ and $\sigmath$ in \eqref{capacity}, we have that
\begin{align}
    \mC & = \mathcal{W} \log_2 \left( 1 + \frac{\mu_0^2 A^2 }{ N_0 \mathcal{W}  }\right). \label{sigmath} 
\end{align}
After substituting the expression of $\mW$ in \eqref{sigmath}, the final expression for channel capacity is
\begin{align}
    \mathcal{C} = \frac{\alpha}{A} \log_2\left( 1 + \frac{\beta_0}{\alpha} A^3  \right), \label{final_capacity}
\end{align}
where 
\begin{align}
\beta_0  \coloneqq \frac{\mu_0^2}{N_0}. \label{beta11}
\end{align}
The factor $\dfrac{\beta_0}{\alpha}A^3$ is the signal-to-noise ratio (SNR) of the optical channel, and the term $\dfrac{\beta_0}{\alpha}$ is the normalized SNR value.




%

{  \subsection{Optimal Photodetector Area}
In order to find the optimal photodetector area that maximizes the channel capacity furnished by \eqref{final_capacity},  we rewrite the capacity in terms of natural log function as
\begin{align}
    \mathcal{C} = \frac{\alpha}{A \ln(2)} \ln\left( 1 + \frac{\beta_0}{\alpha} A^3  \right). \label{final_capacity1}
\end{align}
Taking the partial derivative of \eqref{final_capacity1} with respect to area $A$ and setting it equal to zero, we obtain
\begin{align}
    &\frac{\partial \mC}{\partial A} = 0 \nonumber \\
    &\implies   \left( 1 + \frac{\beta_0}{\alpha} A^3  \right) \ln\left( 1 + \frac{\beta_0}{\alpha} A^3  \right) - 3 \frac{\beta_0}{\alpha} A^3 = 0 \label{e1}
\end{align}
Subtracting three from both sides of \eqref{e1}, we obtain
\begin{align}
   &\left( 1 + \frac{\beta_0}{\alpha} A^3  \right) \ln\left( 1 + \frac{\beta_0}{\alpha} A^3  \right) - 3 \frac{\beta_0}{\alpha} A^3 -3 = -3\nonumber \\
   & \implies \left( 1 + \frac{\beta_0}{\alpha} A^3  \right) \ln\left( 1 + \frac{\beta_0}{\alpha} A^3  \right) - 3 \left(1 + \frac{\beta_0}{\alpha} A^3 \right) = -3 \nonumber \\
   & \implies \left( 1 + \frac{\beta_0}{\alpha} A^3  \right) \left(\ln\left( 1 + \frac{\beta_0}{\alpha} A^3  \right) - 3\right) = -3. \label{e2}
\end{align}
Dividing both sides of \eqref{e2} with $\exp(3)$ and noting that $x = \exp(\ln(x))$,  we have that
\begin{align}
 \left(\ln\left( 1 + \frac{\beta_0}{\alpha} A^3  \right) - 3\right) \exp\left(\ln \left( 1 + \frac{\beta_0}{\alpha} A^3  \right) - 3\right) = -3    \exp(-3). \label{e3}
\end{align}
Now, using the fact that the solution of the equation of the form 
\begin{align}
    x \exp(x) = y \implies x = W(y),
\end{align}
where $W(\cdot)$ is the Lambert W function, we have that \eqref{e3} can be solved as 
\begin{align}
   & \ln\left( 1 + \frac{\beta_0}{\alpha} A^3  \right) - 3 = W\left( -3 \exp(-3) \right) \nonumber \\
    & \implies 1 + \frac{\beta_0}{\alpha} A^3 = \exp\left( W\left( -3 \exp(-3) \right) + 3\right) \nonumber \\
    & \implies A^* = \left( \exp\left( W\left( -3 \exp(-3) \right) + 3\right) - 1 \right)^{\frac{1}{3}} \left( \frac{\alpha}{\beta_0}\right)^{\frac{1}{3}}. \label{opt_A}
\end{align}
Let us define the factor $\gamma_0$ as 
\begin{align}
\gamma_0 \coloneqq \left( \exp\left( W\left( -3 \exp(-3) \right) + 3\right) - 1 \right)^{\frac{1}{3}}.
\end{align} 
To show that the solution \eqref{opt_A} is indeed a maximizer, we evaluate the second derivative of capacity $\mC$ at $A = A^*$. We have that
\begin{align}
    \frac{\partial^2 \mC}{\partial A^2} = \frac{1}{\ln(2)} \left( \frac{-9 \beta_0^2 A^3/\alpha}{\left(1+\frac{\beta_0}{\alpha}A^3 \right)^2} + \frac{2\alpha}{A^3} \ln \left( 1 + \frac{\beta_0}{\alpha} A^3 \right) \right),
\end{align}
and
\begin{align}
  \left.\frac{\partial^2 \mC}{\partial A^2} \right|_{A=A^*} = \frac{\beta_0}{\ln(2)} \left( \frac{-9\gamma_0}{(1+\gamma_0)^2} + \frac{2}{\gamma_0} \ln(1+\gamma_0) \right) \approx -\frac{2.0947\beta_0}{\ln(2)}. 
\end{align}
Since the second derivative is negative at  $A^*$, we have that $A^*$ represents global maximizer of channel capacity. Therefore, from \eqref{opt_A}, we note that the optimal area varies as the cube root of the factor $\frac{\alpha}{\beta_0}$ which is the inverse of normalized SNR value.
}

\section{Ergodic Capacity Maximization Under Intensity Fading in Free-Space Optical Channels} \label{fading_section}

\subsection{Channel Capacity Under Intensity Fading}
The erogodic capacity of a free-space optical channel under fading is described by
\begin{align}
\mC = \int_0^\infty \mathcal{W} \log_2 \left( 1 + \frac{\mu_s^2h^2 }{ N_0 \mathcal{W}  }\right)  \fh\, dh, \label{erg_capacity}
\end{align}   
where $\fh$ is the distribution of random channel coefficient $\mathbbm{h}$, and $\mu_s$ is the maximum signal power captured by the optical receiver under no fading. The signal does not experience fading when the channel coefficient $\mathbbm{h} = 1$ with probability one, i.e., $\mathbbm{P}(\{ \mathbbm{h} = 1 \}) = 1.$ 

\subsection{Close-to-Optimal Detector Area Under Pointing error}\label{analysis:pointing error}
Assuming that the pointing error (due to angle-of-arrival fluctuations) is distributed as a Rayleigh random variable \cite{Bashir:TCOMM:22, Bashir:TWC:22, Tsai:OJCOMS:22}, the received signal power---denoted by $P_s$---captured by the detector is furnished by the relationship:
\begin{align}
    P_s = \mu_s \mathbbm{h}_{\mR} ,
\end{align}
where 
\begin{align}
    \mathbbm{h}_{\mR} \coloneqq \exp \left( - \frac{\mR^2}{2 \rho^2}\right),
\end{align}
where $\mR$ is a {Rayleigh} random variable that models the pointing or misalignment error in the focal plane. The probability density function (pdf) of $\mR$ is defined as 
\begin{align}
f_\mR(r) \coloneqq \dfrac{r}{\sigma_p^2} \exp\left(-\dfrac{r^2}{2\sigma_p^2} \right) \cdot \mathbf{1}_{[0, \infty)}(r),
\end{align}
where $\sigma_p$ is the \emph{scale parameter} of the pdf $f_\mR(r)$. By noting that $h_\mR$ is the realization of $\mathbbm{h}_{\mR}$, the ergodic capacity---under pointing error---is given by
\begin{align}
\mC &= \int_0^\infty \mathcal{W} \log_2 \left( 1 + \frac{\mu_s^2 h_{\mR}^2 }{ N_0 \mathcal{W}  }\right)  f_\mR(r)\, dr, \\
  & = \int_0^\infty \mathcal{W} \log_2 \left( 1 + \frac{\mu_s^2 \exp \left( - \frac{r^2}{ \rho^2}\right)  }{ N_0 \mathcal{W}  }\right) \dfrac{r}{\sigma_p^2} \exp\left(-\dfrac{r^2}{2\sigma_p^2} \right)  \, dr \\ 
  & = \int_0^\infty \frac{\alpha}{A} \log_2 \left( 1 + \frac{\mu_0^2  \exp \left( - \frac{r^2}{ \rho^2}\right)  }{ N_0 \alpha  } A^3\right) \dfrac{r}{\sigma_p^2} \exp\left(-\dfrac{r^2}{2\sigma_p^2} \right)  \, dr \\
  & = \int_0^\infty \frac{\alpha}{A} \log_2 \left( 1 + \frac{ \beta(r)  }{  \alpha  } A^3\right) \dfrac{r}{\sigma_p^2} \exp\left(-\dfrac{r^2}{2\sigma_p^2} \right)  \, dr, 
  \label{capacity_pointing}
\end{align}
where 
\begin{align}
    \beta(r) \coloneqq \frac{\mu_0^2  \exp \left( - \frac{r^2}{ \rho^2}\right)  }{ N_0 }, r > 0.
\end{align}
 Let us denote the optimal photodetector area $A^*$ for a realization $r$ of random variable $\mR$ as $A^*(r)$ where 
 \begin{align}
 A^*(r) = \gamma_0 \left(\frac{\alpha}{\beta(r)} \right)^{\frac{1}{3}} = \gamma_0 \left(\frac{\alpha N_0}{ \mu_0^2  \exp \left( - \frac{r^2}{ \rho^2}\right) } \right)^{\frac{1}{3}}.
 \end{align}
  The (suboptimal) average  area in this case is
 \begin{align}
     &A^\star = \int_0^\infty A^*(r)  \dfrac{r}{\sigma_p^2} \exp\left(-\dfrac{r^2}{2\sigma_p^2} \right)  \, dr \\
     & = \int_0^\infty \gamma_0 \left(\frac{\alpha N_0}{ \mu_0^2  \exp \left( - \frac{r^2}{ \rho^2}\right) } \right)^{\frac{1}{3}}  \dfrac{r}{\sigma_p^2} \exp\left(-\dfrac{r^2}{2\sigma_p^2} \right)  \, dr \nonumber \\
     & = \gamma_0 \left( \frac{\alpha N_0}{\mu_0^2} \right)^{\frac{1}{3}} \int_0^\infty \frac{r}{\sigma_p^2} \exp\left( - r^2 \left( \frac{1}{2\sigma_p^2} - \frac{1}{3\rho^2} \right) \right) \, dr, \label{integral1}
 \end{align}
where the integral in \eqref{integral1} converges if the factor 
\begin{align}
&\frac{1}{2\sigma_p^2} - \frac{1}{3\rho^2} > 0  \implies \rho > \sqrt{\frac{2}{3}} \sigma_p. \label{cond1}
\end{align}
This condition implies that the spot radius has to be at least $\sqrt{\frac{2}{3}}$ times the pointing error scale factor $\sigma_p$. This condition is almost always satisfied since the beam footprint is made large enough to accommodate any deviations due to misalignment or pointing error. In most scenarios, a suboptimal value of beam radius $\rho$ (in relation to $\sigma_p$) is given by $\rho = \sqrt{2} \sigma_p$ \cite{Bashir:TWC:22}.

When the condition \eqref{cond1} holds, we have that
\begin{align}
   &A^\star = \gamma_0 \left( \frac{\alpha N_0}{\mu_0^2} \right)^{\frac{1}{3}} \int_0^\infty  \frac{r}{\sigma_p^2} \exp\left( - r^2 \left( \frac{3\rho^2-2\sigma_p^2}{6 \rho^2\sigma_p^2}  \right) \right) \, dr \nonumber \\
   & = \gamma_0 \left( \frac{\alpha N_0}{\mu_0^2} \right)^{\frac{1}{3}} \int_0^\infty  \frac{r}{\sigma_p^2} \exp\left( -  \left( \frac{r^2}{ 2\left(\frac{ 3 \rho^2\sigma_p^2}{3\rho^2-2\sigma_p^2}\right)}  \right) \right) \, dr. \label{integral2}
\end{align}
Dividing and multiplying \eqref{integral2} by the factor $\dfrac{3\rho^2}{3\rho^2-2\sigma_p^2}$, we obtain the suboptimal photodetector area as
\begin{align}
A^\star & = \gamma_0 \left( \frac{\alpha N_0}{\mu_0^2} \right)^{\frac{1}{3}} \left(\frac{3\rho^2}{3\rho^2-2\sigma_p^2} \right)  \int_0^\infty  \frac{r}{\frac{3\rho^2\sigma_p^2}{3\rho^2-2\sigma_p^2}} \exp\left( -  \left( \frac{r^2}{ 2\left(\frac{ 3 \rho^2\sigma_p^2}{3\rho^2-2\sigma_p^2}\right)}  \right) \right) \, dr \label{int3} \\
& = \gamma_0 \left( \frac{\alpha N_0}{\mu_0^2} \right)^{\frac{1}{3}} \left(\frac{3\rho^2}{3\rho^2-2\sigma_p^2} \right), \label{pointing_error_soln}
\end{align}
since the integral in \eqref{int3} evaluates to unity.

\subsection{Suboptimal Detector Area Under Strong Atmospheric Turbulence}
Assuming the case of strong atmospheric turbulence that is modeled by the negative exponential distribution, we note that the suboptimal average area in this case is
\begin{align}
    A^\star &= \int_0^\infty A^*(h) \frac{1}{\eta} \exp\left(-\frac{1}{\eta} h \right) \, dh, \nonumber\\
    &= \int_0^\infty \gamma_0 \left( \frac{\alpha}{\beta(h)} \right)^{\frac{1}{3}} \frac{1}{\eta} \exp\left(-\frac{1}{\eta} h \right) \, dh,
\end{align}
where $\eta > 0$ is the mean of negative exponential distribution and $\beta(h) = \dfrac{(\mu_0 h)^2}{N_0}$. This implies that
\begin{align}
    A^\star &= \frac{ \gamma_0 \alpha^{\frac{1}{3}} N_0^{\frac{1}{3}}}{\eta \mu_0^{\frac{2}{3}}} \int_0^\infty h^{-\frac{2}{3}}  \exp\left(-\frac{1}{\eta} h \right)\, dh \nonumber \\
    & = \frac{ \gamma_0 \alpha^{\frac{1}{3}} N_0^{\frac{1}{3}}}{\eta \mu_0^{\frac{2}{3}}} \eta^{\frac{1}{3}}\int_0^\infty t^{-\frac{2}{3}} \exp(-t) \, dt \, dt \nonumber \\
    & = \frac{ \gamma_0 \alpha^{\frac{1}{3}} N_0^{\frac{1}{3}} \Gamma(\frac{1}{3})}{\mu_0^{\frac{2}{3}} \eta^{\frac{2}{3}}}, \label{exp_turbulence_soln}
\end{align}
where the gamma function $\Gamma(\cdot)$ is defined as $\Gamma(z) \coloneqq \int_0^\infty t^{z-1} \exp(-t) \, dt.$

\section{Channel Capacity Maximization of Detector Array Receivers } \label{detector array}
\begin{figure}
    \centering
    \includegraphics[scale=1.2]{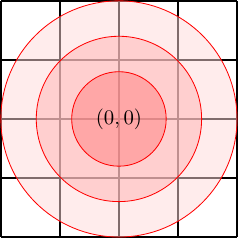}
    \caption{Gaussian beam projected onto a $4\times 4$ detector array.}
    \label{Fig2}
\end{figure}

In this section, we consider the channel capacity of a detector array receiver in free-space optical communications. Detector arrays are commonly used for tracking the angle-of-arrival of laser beams by virtue of estimation of spot position on a detector array in focal plane \cite{Bashir:TCOMM:a:21, Bashir:TAES:16}. However, detector arrays can also be used in a joint beam tracking/symbol detection configuration which simplifies receiver hardware by getting rid of a separate beam tracking assembly dedicated exclusively to beam tracking\cite{Tsai:OJCOMS:22}.  

As shown in Fig.~\ref{Fig2}, the detector array considered in this study is of a square configuration where the individual elements or detectors are also of a square shape and are arranged in an $n \times n$ arrangement for $n$  a positive integer. Therefore, the total number of individual detectors in the array is $M = n^2$.  We further assume that the area of the array is $\mathcal{A}$ m$^2$. Without any loss of generality, we assume that the array is centered at the origin $(0, 0)$.

For thermal noise limited regime, the output $Y_m$ of the $m$th element of the detector array is given by
\begin{align}
    Y_m = \Lambda_m + X_m, \quad m = 1, 2, \dotsc, M, \label{Ym}
\end{align}
In \eqref{Ym}, the signal component $\Lambda_m$ is defined as
\begin{align}
    \Lambda_m \coloneqq \iint_{A_m} \Lambda_s(x, y) \, dy \, dx, \label{lambda_m}
\end{align}
where $\Lambda_s$ is furnished by \eqref{intensity_def}. The noise component $X_m$ is distributed as a Gaussian random varaible: $\mathcal{N}(0, \sigma_{\textsf{th}}^2=N_0W)$. We note that $X_i$ and $X_j$ are independent and identically distributed random variables for integers $i$ and $j$ such that $i \neq j$ and $1 \leq i, j \leq M$.

In order to form the sufficient statistic with an array of detectors, we look at two signal combining schemes: the \emph{equal gain combiner} (EGC) and the \emph{maximal ratio combiner} (MRC). The EGC combines the output of each detector of the array with uniform weights, whereas the MRC weights the output of each detector with the optimal intensity value before combining \cite{Tsai:OJCOMS:22, Bashir:TAES:18}. It is important to note here that the intensity in each detector is not the same as the intensity in every other detector since the intensity of the laser spot in the focal plane varies according to Gaussian distribution as given by \eqref{intensity_def}. In the presence of random fluctuation of angle-of-arrival, the spot location on the array will change randomly. In such a scenario, diversity schemes such as the MRC are quite effective in countering the effect of misalignment  provided that the spot location on the array can be tracked effectively in real time \cite{Bashir:TCOMM:a:21, Tsai:OJCOMS:22}. Here, we also remind the reader that the MRC algorithm, discussed in this paper, does not take into account the phase of the optical signal since we are dealing with intensity modulated/direct detection (IM/DD) scheme in this study.

In the upcoming sections, we look at the ECG and MRC combining schemes for a detector array receiver. We will also consider  the optimization of individual detector area of the array that maximizes the capacity that is achieved with these combining schemes.

\subsection{Equal Gain Combining}
An equal gain combiner simply adds output from each element of the array to form the sufficient statistic $Y$ as follows:
\begin{align}
    Y &= \sum_{m=1}^M Y_m =  S + X,  
\end{align}
where $S = \sum_{m=1}^M \Lambda_m$ is the signal component  and $X = \sum_{m=1}^M X_m$ is the noise component. Since the noise in each detector of the array is assumed to be independent, the variance of $X$ is $M N_0 W$. The channel capacity of this scheme is
\begin{align}
    \mC & = \mW \log_2\left(1 + \frac{\left(\sum_{m=1}^M \Lambda_m\right)^2}{M N_0 \mW} \right) \\
    & = \mW \log_2\left(1 + \frac{\mu_s^2}{M N_0 \mW} \right)
    \label{capacity_EGC}
\end{align}
where $\mu_s \coloneqq \sum_{m=1}^M \Lambda_m$ is the total energy captured by the detector array.  The bandwidth $\mW$ in \eqref{capacity_EGC} is a function of number of detectors $M$ in the array provided the array area $\mA$ is fixed. It is given by 
\begin{align}
    \mW = \frac{d}{2 \pi \epsilon_0 \epsilon_r R A} = \frac{d M}{2 \pi \epsilon_0 \epsilon_r R \mathcal{A}}, \label{BW_Combining}
\end{align}
where $A$ is the area of a single detector in the array.

\subsection{Maximal Ratio Combining}
The output $Y_m$ of the $m$th element of the array is weighted by the signal component $\Lambda_m$ before fusion takes place for the formation of the sufficient statistic. The resulting sufficient statistic, $Y$, is
\begin{align}
    Y = \sum_{m=1}^M Y_m \Lambda_m = S + X,
\end{align}
where the signal component $S = \sum_{m=1}^M \Lambda_m^2$ and the noise component $X = \sum_{m=1}^M \Lambda_m X_m$. We note that $X$ is a zero-mean Gaussian random variable with variance $\sigma_X^2 =N_0 W \sum_{m=1}^M \Lambda_m^2$. The channel capacity of this scheme is
\begin{align}
    \mathcal{C} &= \mW \log_2 \left( 1 + \frac{S^2}{\sigma_X^2} \right) =\mW \log_2 \left( 1 + \frac{\left( \sum_{m=1}^M \Lambda_m^2 \right)^2}{N_0 \mW \sum_{m=1}^M \Lambda_m^2} \right) \nonumber \\
    & = \mW \log_2 \left( 1 + \frac{ \sum_{m=1}^M \Lambda_m^2 }{N_0 \mW  } \right), \label{capacity_MRC}
\end{align}
where $\mW$ is given by \eqref{BW_Combining}.

\subsection{Optimization Problem}
 For maximization of channel capacity with an array of detectors, we optimize the area $A$ of the single detector element while we keep the total area of the array constant. Thus, we consider the optimization problem of the following type:
\begin{equation}
\begin{aligned}
\underset{A}{\text{maximize}} \quad & \mathcal{C}\\
\textrm{subject to} \quad &  \, \mathcal{A} = \mathcal{A}_0. \label{opt1}
\end{aligned}
\end{equation}
In this optimization problem, $\mathcal{C}$ is furnished by \eqref{capacity_EGC} for the equal gain combiner and  \eqref{capacity_MRC} for maximal ratio combining. The factor $\mathcal{A}_0$ is a  positive constant.

\subsubsection{Optimization of Photodetector Area for EGC Array Receiver}

In order to obtain a closed-form solution to \eqref{opt1} for the EGC array, we rewrite \eqref{capacity_EGC} as
\begin{align}
    \mC & =\mW \log_2\left(1 + \frac{\mu_s^2}{M N_0 \mW} \right) \\
    & = \frac{\alpha}{A} \log_2\left(1 + \frac{\mu_s^2}{\mA N_0 \alpha} A^2 \right),
\end{align}   
where we used the fact that the number of detectors $M = \frac{\mA}{A}$, and that the bandwidth of the array is the same as the bandwidth of each individual element or detector of the array: $\mW = \frac{\alpha}{A}$. Let us denote the factor
\begin{align}
    \beta_1 \coloneqq \frac{\mu_s^2}{N_0} = \frac{\left(\sum_{m=1}^M \Lambda_m\right)^2}{N_0}.
\end{align}
Then, 
\begin{align}
    \mC &= \frac{\alpha}{A} \log_2\left(1 + \frac{\beta_1}{\mA  \alpha} A^2 \right) \\
    &= \frac{\alpha}{A\ln(2)} \ln\left(1 + \frac{\beta_1}{\mA  \alpha} A^2 \right). \label{egc_final_capacity}
\end{align}
The quantity $\frac{\beta_1}{\mA \alpha}$ is the normalized SNR of the EGC array. Here, we note that when the individual detector area $A$ is fixed, increasing array area $\mA$ implies increasing the number of detectors $M$ in the array. Thus, for fixed $A$, increasing $\mA$ implies increasing the total noise in the array since noise power increases linearly with the number of elements $M$ of the array.

In order to find the maximum of $\mC$ in \eqref{egc_final_capacity} as a function of area $A$, we take the derivative of $\mC$ as a function of $A$ and set it equal to zero to obtain
\begin{align}
  \left(1 + \frac{\beta_1}{\mA  \alpha} A^2 \right) \ln\left(1 + \frac{\beta_1}{\mA  \alpha} A^2 \right) - 2\frac{\beta_1}{\mA \alpha}A^2 = 0. \label{eq4}  
\end{align}
Subtracting two from both sides of \eqref{eq4}, we obtain
\begin{align}
\left(1 + \frac{\beta_1}{\mA  \alpha} A^2 \right) \left( \ln\left(1 + \frac{\beta_1}{\mA  \alpha} A^2 \right) - 2 \right) = -2. \label{eq5}
\end{align}
Representing $\left(1 + \frac{\beta_1}{\mA  \alpha} A^2 \right)$ by $\exp\left( \ln \left(1 + \frac{\beta_1}{\mA  \alpha} A^2 \right) \right)$
and dividing both sides of \eqref{eq5} by $\exp(2)$, we have that
\begin{align}
  &\exp\left( \ln \left(1 + \frac{\beta_1}{\mA  \alpha} A^2 \right) -2 \right)  \left( \ln\left(1 + \frac{\beta_1}{\mA  \alpha} A^2 \right) - 2 \right) = -2\exp(-2) \\
  &\implies  \ln \left(1 + \frac{\beta_1}{\mA  \alpha} A^2 \right) -2  = W \left( -2 \exp(-2) \right) \\
  & \implies 1 + \frac{\beta_1}{\mA  \alpha} A^2 = \exp \left( W \left( -2 \exp(-2) \right) + 2 \right) \\
  & \implies A^* =   \left(\exp \left( W \left( -2 \exp(-2) \right) + 2 \right) - 1 \right)^{\frac{1}{2}} \left( \frac{\alpha \mA}{\beta_1} \right)^{\frac{1}{2}}. \label{opt_A_egc}
\end{align}
Let us define $\gamma_1 \coloneqq \left(\exp \left( W \left( -2 \exp(-2) \right) + 2 \right) - 1 \right)^{\frac{1}{2}}$. In order to show that $A^*$ is indeed a global maximizer, we evaluate the second partial of channel capacity at  $A=A^*$:
\begin{align}
   \left. \frac{\partial^2 \mC}{\partial A^2}\right|_{A=A^*} &= \frac{\beta_1^{\frac{3}{2}}}{\alpha^{\frac{1}{2}} \mA^{\frac{3}{2}} \ln(2)} \left( -\frac{4\gamma_1}{\left(1+\gamma_1^2\right)^2} - \frac{2}{\gamma_1\left(1+\gamma_1^2 \right)} + \frac{2 \ln\left(1+\gamma_1^2 \right)}{\gamma_1^3} \right) \\
   & \approx - \frac{0.1218\beta_1^{\frac{3}{2}}}{\alpha^{\frac{1}{2}} \mA^{\frac{3}{2}} \ln(2)}
\end{align}
which is strictly negative at $A=A^*$. Hence, $A^*$ in \eqref{opt_A_egc} is indeed a global maximizer of the channel capacity achieved with an EGC array. 

From \eqref{opt_A_egc}, we observe that the optimal area is proportional to the square root of the inverse of normalized SNR which is in contrast to the cube root relationship for a single-detector optical receiver (see \eqref{opt_A}).

\subsubsection{Optimization of Photodetector Area for MRC Array Receiver}
In order to maximize capacity of an MRC array---given by \eqref{capacity_MRC}---as a function of photodetector area, we first define the average signal intensity $\lambdasm$ (measured in Watts/unit area) on the $m$th detector of the array:
\begin{align}
    \lambdasm \coloneqq \frac{\Lambda_m}{A}.
\end{align}
We now rewrite \eqref{capacity_MRC} as a function of $\lambdasm$ as
\begin{align}
   \mC &= \mW \log_2 \left( 1 + \frac{ A^2 \sum_{m=1}^M \left(\lambdasm \right)^2 }{N_0 \mW  } \right) \\
   & = \frac{\alpha}{A} \log_2 \left( 1 + \frac{\beta_2}{\alpha}A^3 \right)
\end{align}
where $\beta_2 \coloneqq \dfrac{\sum_{m=1}^M \left(\lambdasm \right)^2}{N_0}$. We note that the channel capacity of an MRC array has the same form as the channel capacity of a single detector (given by \eqref{final_capacity}). Therefore, the optimum detector area for the MRC array is 
\begin{align}
   A^* = \left( \exp\left( W\left( -3 \exp(-3) \right) + 3\right) - 1 \right)^{\frac{1}{3}} \left( \frac{\alpha}{\beta_2}\right)^{\frac{1}{3}}. \label{opt_A_mrc} 
\end{align}
Here, the optimal photodetector area varies as cube root of normalized SNR as opposed to square root relationship for EGC array.

\section{Channel Capacity Under Signal Dependent Shot Noise} \label{analysis: general case}
In this section, we consider the channel capacity maximization for the general case where the channel suffers both from shot noise and thermal noise. Let the  number of received signal photons be equal to $N_s$. Then, $\E[N_s] = \mu_s$ where $\E[\cdot]$ is the ensemble average operator. The average background radiation intensity (giving rise to dark current) is denoted by $\lambda_b$ Watts/m$^2$. The total noise power due to background radiation is $\mu_b \coloneqq \lambda_b A$. Let the average number of noise photons be $N_b$. In this case, $\E[N_b] = \mu_b$. 

Here, for the sake of simplicity, let us consider the popular on-off keying (OOK) signaling. For the ``on'' state,  let the total number of photons (signals plus noise ) be denoted by $N_t = N_s + N_b$. In low-to-medium photon rate regime, the random variable $N_t$ is modeled by a Poisson distribution with parameter $(\mu_s + \mu_b)$. In high photon rate regime, $N_t$ converges (in distribution) to a Gaussian random variable: $N_t \sim \mathcal{N}(\mu=\mu_s + \mu_b, \sigmash^2=\mu_s + \mu_b)$ when the signal pulse is present. When no signal is present, $N_t \sim \mathcal{N}(\mu= \mu_b, \sigmash^2 = \mu_b)$. Here, $\sigmash^2$ represents the (signal dependent) power in \emph{shot noise}.

In the high photon rate regime, the signal-to-noise ratio (denoted by the symbol $\SNR$) is given by
\begin{align}
    \SNR & = \frac{\mu_s^2}{\sigma_{\textsf{sh}}^2 + \sigmath^2}= \frac{\mu_s^2}{\mu_s + \mu_b + \sigmath^2} \nonumber \\
    & \overset{A \ll \pi \rho^2}{\approx} \frac{\mu_0^2 A^2}{\mu_0 A + \lambda_b A + \sigmath^2} = \frac{\mu_0^2 A }{\mu_0 + \lambda_b + \frac{\sigmath^2}{A}}.
\end{align}
Thus, the $\SNR \to 0$ as $A \to 0$. 

 For the binary OOK channel, the output $Y$ is related to random input $X$ as
\begin{align}
    Y = X + N_\textsf{sh} + N_{\textsf{th}},
\end{align}
where the random variable $N_{\textsf{sh}}$ corresponds to signal dependent shot noise with distribution $\mathcal{N}(0, \sigma_{\textsf{sh}}^2)$, and $N_{\textsf{th}}$ represents thermal noise with distribution $\mathcal{N}(0, \sigma_{\textsf{th}}^2)$. We assume that $N_\textsf{sh} \perp N_{\textsf{th}}$. For this channel, we obtain the capacity as the maximization of the difference of entropy $H(X)$ and conditional entropy $H(X|Y)$:
\begin{align}
    \mC_0 \coloneqq  \max_{p(x)} \left(H(X) - H(X|Y) \right)  \label{capacity1}
\end{align}
where $\mC_0$ is capacity in bits per channel use and $p(x)$ is the distribution of input $X$. The capacity $\mC$ in bits per second is given by 
\begin{align}
    \mC = W \mC_0.  \label{capacity_shot_noise}
\end{align}
  In \eqref{capacity1}, the input $X$ is a discrete random variable and the output $Y$ is sampled from a continuous distribution. The quantity $H(X) \coloneqq - \sum_{x \in \mathcal{X}} p(x) \log_2 p(x),$ is the entropy of the input $X$ where the set $\mathcal{X} \coloneqq  \{0, 1\}$. The conditional entropy is
\begin{align}
    H(X|Y) & \coloneqq  -\sum_{x \in \mathcal{X}} \int_y p(x) p(y|x) \log_2  \frac{p(y|x) p(x)}{p(y)} \, dy, 
\end{align}
where $p(y) \coloneqq  \sum \limits_{x \in \mathcal{X}} p(y|x = i) p\left(  x =i \right) $. Moreover, 
\begin{align}
    p(y|x=1) &= \frac{1}{\sqrt{2 \pi (\mu_0 A + \lambda_b A + \sigmath^2)}} \nonumber  \exp\left( - \frac{(y - \mu_0 A - \lambda_bA)^2}{2 (\mu_0A + \lambda_b A + \sigmath^2)} \right), \\
    p(y|x=0) &= \frac{1}{\sqrt{2 \pi ( \lambda_b A + \sigmath^2)}} \exp\left( - \frac{(y -  \lambda_bA)^2}{2 ( \lambda_bA + \sigmath^2)} \right).
\end{align}
The closed-form solution of optimal detector area that maximizes channel capacity in \eqref{capacity_shot_noise} is not easy to achieve. Therefore, for signal dependent shot noise scenario, we resort to finding the peak of channel capacity numerically (graphically) in Section~\ref{experiments}.

\section{Experimental Results and Discussion}\label{experiments}
In this section, we explain and interpret the experimental results obtained for this study. The default parameter values are highlighted in Table.~\ref{default_params}.
\begin{table}[H]
    \centering
    \caption{Default parameter values for simulations}
    \label{default_params}
    \scalebox{0.95}{
    \begin{tabular}{|l|c|c|}
    \hline
    Parameter & Symbol& Default value \\
    \hline 
        Diode junction thickness &$d$ &  0.1 $\mu$m\\
        Permittivity (free space)&$\epsilon_0$ & $8.854 \times 10^{-12}$ F/m \\
        Permittivity (relative)&$\epsilon_r$ & 12.95 \\
        Noise PSD &$N_0$ & $4.11\times 10^{-21}$ W/Hz\\
        Photodiode resistance &$R$ & 10 ohms \\
        Peak received intensity &$\mu_0$ & 10 mW \\
        Background radiation intensity &$\lambda_b$ & $\frac{\mu_0}{20}$ to $\frac{\mu_0}{3}$  \\
        Focal spot radius &$\rho$ & 2 mm \\
        Number of detectors in the array &$M$ & 16 \\
        Photodetector area &$A$ & $10^{-9}$ m$^2$\\
        Photodetector array area &$\mathcal{A}$ & 4 mm$^2$\\
        Pointing error scale factor & $\sigma_p$& 0.01--8 mm\\
        Exponential fading distribution mean & $\eta$ & 0.1--0.8\\
         \hline 
    \end{tabular}
    }
    \end{table}

\begin{figure}
    \centering
    \includegraphics[scale=0.62]{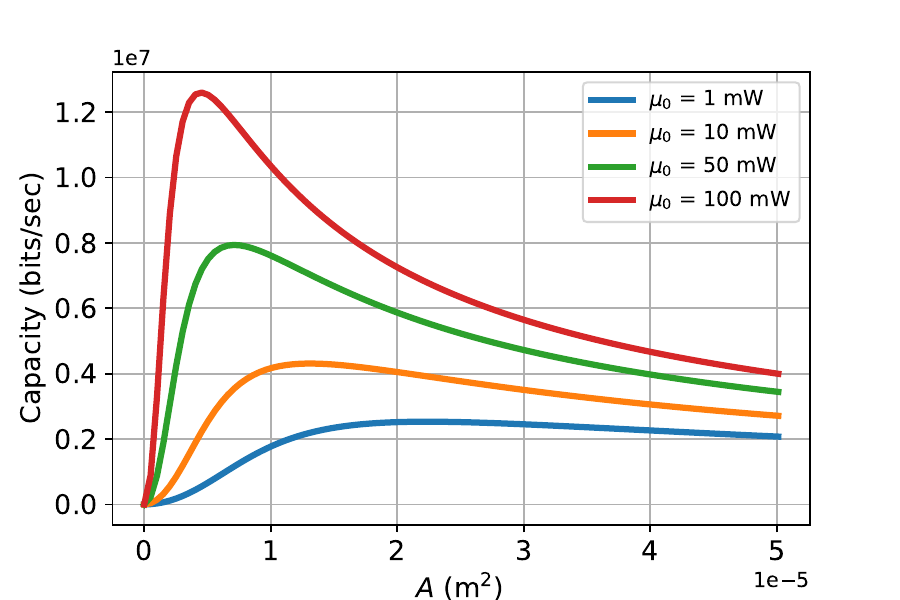}
    \caption{Channel capacity for thermal noise dominated regime.}
    \label{fig1}
\end{figure}

Fig.~\ref{fig1} depicts the channel capacity as a function of photodetector area for the thermal noise limited regime where the noise statistics  is assumed to follow the Gaussian distribution. This figure plots the channel capacity as a function of photodiode area $A$ for different values of received signal power $\mu_0$. We observe that the channel capacity is maximized at a certain (optimal) value of photodiode area. We also note from this figure that as the received signal power $\mu_0$ is increased, the optimal area shifts to the left towards zero. This is explained by the inverse dependence of the optimal photodetector area on the signal-to-noise ratio as highlighted by \eqref{opt_A}. 


\begin{figure}
    \centering
    \includegraphics[scale=0.62]{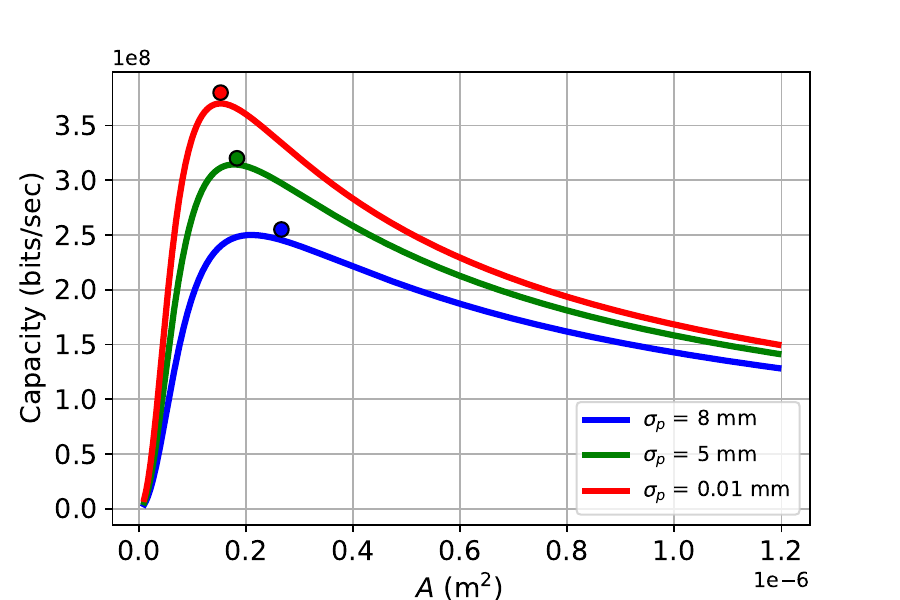}
    \caption{Channel capacity as a function of pointing error standard deviation $\sigma_p$.}
    \label{fig4}
\end{figure}

\begin{figure}
    \centering
    \includegraphics[scale=0.62]{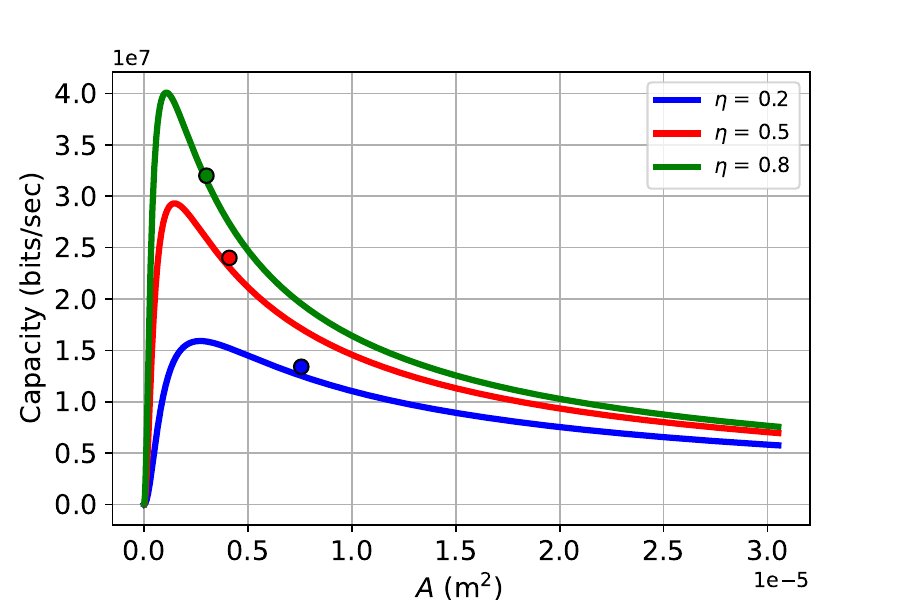}
    \caption{Channel capacity maximization with respect to photodetector area under atmospheric turbulence conditions.}
    \label{fig8}
\end{figure}

Fig.~\ref{fig4} depicts channel capacity for different values of pointing error (modeled by Rayleigh distribution) scale factor $\sigma_p$. Here, we observe that the optimal photodiode area depends on the value of $\sigma_p$. A larger value of $\sigma_p$ minimizes the average signal-to-noise ratio, and this fact implies that the optimal area should shift to the right to compensate for the reduced signal-to-noise ratio when the scale factor $\sigma_p$ is large.  A related figure, Fig.~\ref{fig8} depicts channel capacity for signal fading due to strong  turbulence that is modeled by the negative exponential fading. These two figures are also compared with the suboptimal solutions presented by \eqref{pointing_error_soln} for the channel suffering from  pointing error exclusively, and \eqref{exp_turbulence_soln} for strong turbulence condition. The suboptimal solutions are shown by circular markers of the same color as the capacity curves. We note that the suboptimal solutions are quite close to the actual optimal solution (the area value right below the peak capacity) especially for the pointing error case. We also note that the approximation with suboptimal solution gets better with increasing SNR at the receiver. 

\begin{figure}
    \centering
    \includegraphics[scale=0.62]{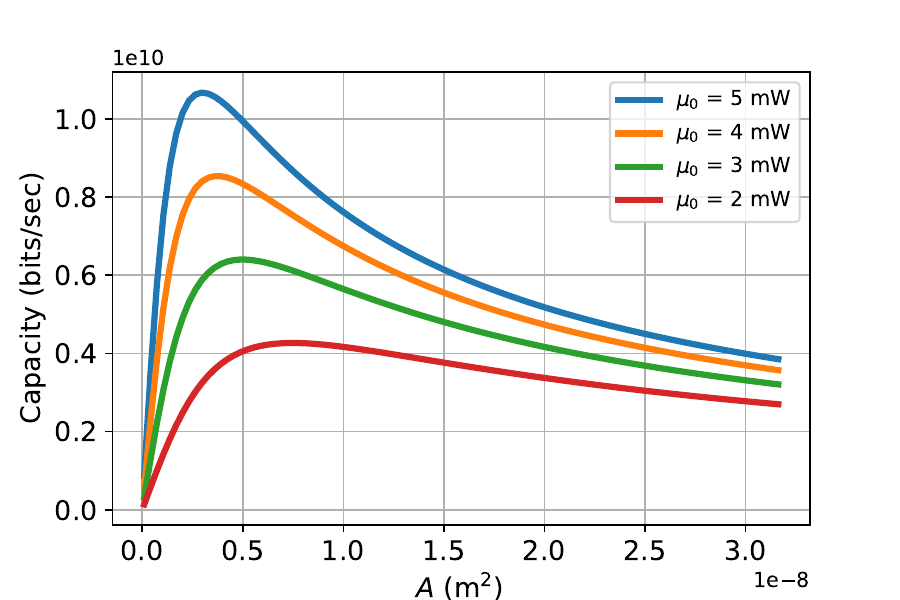}
    \caption{Channel capacity as a function of element or detector area $A$ for EGC array.}
    \label{fig3}
\end{figure}

\begin{figure}
    \centering
    \includegraphics[scale=0.62]{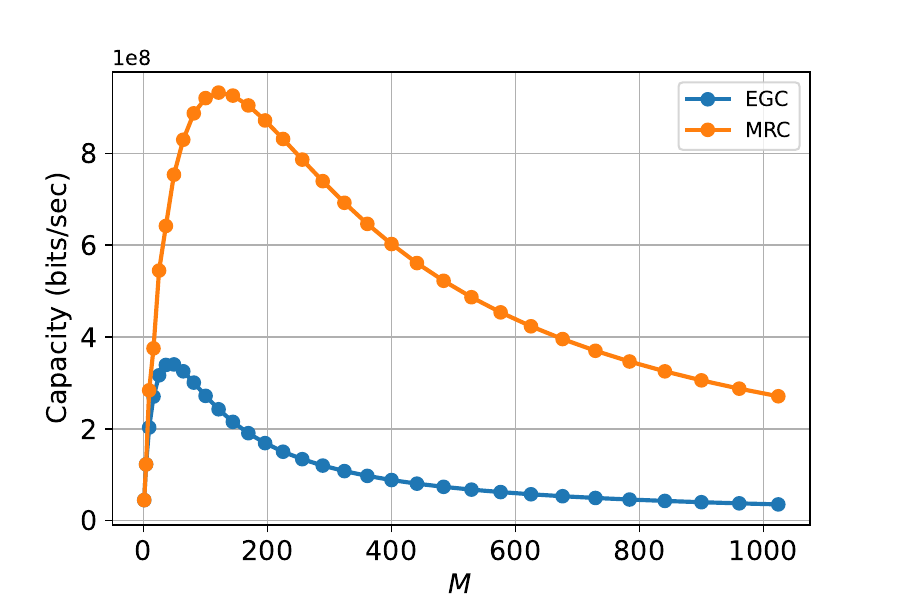}
    \caption{Channel capacity as a function of number of detectors $M$ in the array for MRC and EGC combining techniques.}
    \label{fig7}
\end{figure}

For optical receivers based on detector arrays, Fig.~\ref{fig3} and Fig.~\ref{fig7} correspond to the channel capacity maximization problems for EGC and MRC arrays. Here,  we have assumed a square array with square individual detectors as shown in Fig.\ref{Fig2}. Fig.~\ref{fig3} shows the channel capacity plots for the EGC array receiver as a function of individual detector area $A$ when the array area $\mA$ is fixed at 4 square millimeters. In Fig.~\ref{fig7}, the channel capacity  is plotted as a function of total number of detectors $M$ in the array for the EGC and MRC arrays for a fixed array area $\mA$. Here, for a particular value of $M$, the area of each detector is $A=\frac{\mathcal{A}}{M} = \frac{{4 \text{mm}^2}}{M}$. We note that there is a certain number of elements $M^*$ (or $A^*$) at which the channel capacity is maximized. Moreover, since the MRC minimizes the effect of noise---or alternatively, maximizes the signal-to-noise ratio---the optimal area $A^*$ shifts to the left (relative to optimal area for EGC) in order to maximize the bandwidth and the resulting channel capacity. Since $\mathcal{A}$ is constant, this implies that  the optimal $M^*$ should shift to the right because $M^* = \frac{\mathcal{A}}{A^*}$. We also note that the channel capacity degrades if the number of detectors or elements in the array is too large. This is due to the fact that each detector contributes thermal noise  because of the resistance generated by detector contacts and circuit wires, and the variance of this noise is independent of the area of each detector. If the number of detectors is two large, the noise contribution grows linearly with the number of detectors and the resulting signal-to-noise and capacity degrades significantly.

\begin{figure}
    \centering
    \includegraphics[scale=0.62]{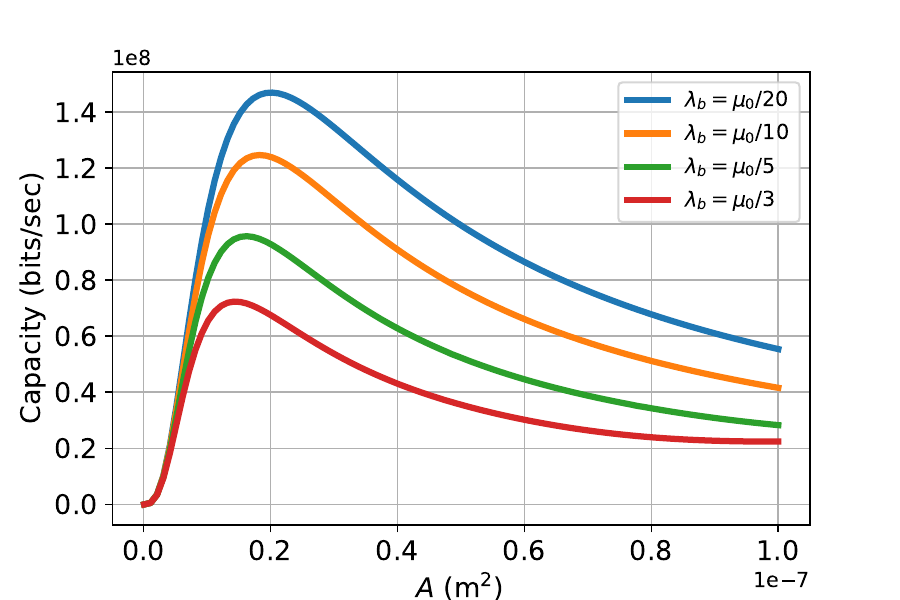}
    \caption{Channel capacity for background radiation dominated shot noise regime.}
    \label{fig2}
\end{figure}

Fig.~\ref{fig2} shows the channel capacity for the general noise (Section~\ref{analysis: general case}) case where the total noise is the sum of  both the signal dependent and signal independent Gaussian noise components. Here, the capacity curves are plotted for different values of background radiation power $\lambda_b$, and for this set of curves, we assume that the shot noise generated by background radiation dominates thermal noise. Here, we observe that the channel capacity decreases monotonically with $\lambda_b$. We also note that the optimal area decreases (shifts to the left) with increasing value of $\lambda_b$. This is due to the fact that the total background noise power---$\lambda_b A$---is a function of area $A$. In order to minimize the effect of large $\lambda_b$, the area has to shrink in order to maximize channel capacity. This is in contrast to the thermal noise limited regime case in Fig.~\ref{fig1} where the noise power was independent of detector area $A$, and the optimal area shifted to the right with increasing noise power. 

\section{Conclusion} \label{conclusion}
In this paper, we carried out a detailed analysis of the relationship between the photodetector area and the capacity of a free-space optical channel. For the signal independent Gaussian noise optical channel, we discovered that the  analytical solutions for optimal area are functions of the term $\left( \frac{\alpha}{\beta} \right)^{\frac{1}{m}}$ where $\frac{\beta}{\alpha}$ is the normalized SNR.  This result indicates that the optimal photodetector area is inversely proportional to signal-to-noise ratio at the receiver. We also carried out channel capacity maximization for intensity fading channels and derived suboptimal solutions that provide quite a good approximation to optimal solutions, especially at high SNR. For the signal dependent shot noise channel that is dominated by noise from background radiation, we discovered that  the optimal area increases with increasing SNR at the receiver, a fact that is opposite to the channel capacity behavior for signal independent Gaussian noise. The results of this study can be used to devise a novel optical wireless receiver that comprises a number of photodetectors of different areas (see Fig.~\ref{adaptive_receiver}). For such a receiver, a photodetector of  appropriate area can be chosen---based on the estimated value of signal intensity---in order to mitigate the effect of slow intensity fading either due to atmospheric turbulence or pointing error.

\appendices

\bibliography{references.bib}
\bibliographystyle{ieeetr}

\end{document}